\documentclass[pre,aps,10pt,twocolumn,superscriptaddress,longbibliography,floatfix,dvipsnames]{revtex4-2}

\usepackage[colorlinks=true,linkcolor=blue,urlcolor=blue,citecolor=blue,breaklinks=true]{hyperref}
\usepackage{amsmath,amsthm,amssymb} 

\usepackage{graphicx}
\usepackage[dvipsnames]{xcolor}

\begin{document}

\title{Discontinuous stochastic forcing in Greenland ice core data}

\author{Keno Riechers\,\href{https://orcid.org/0000-0002-1035-9960}{\includegraphics[width=3.2mm]{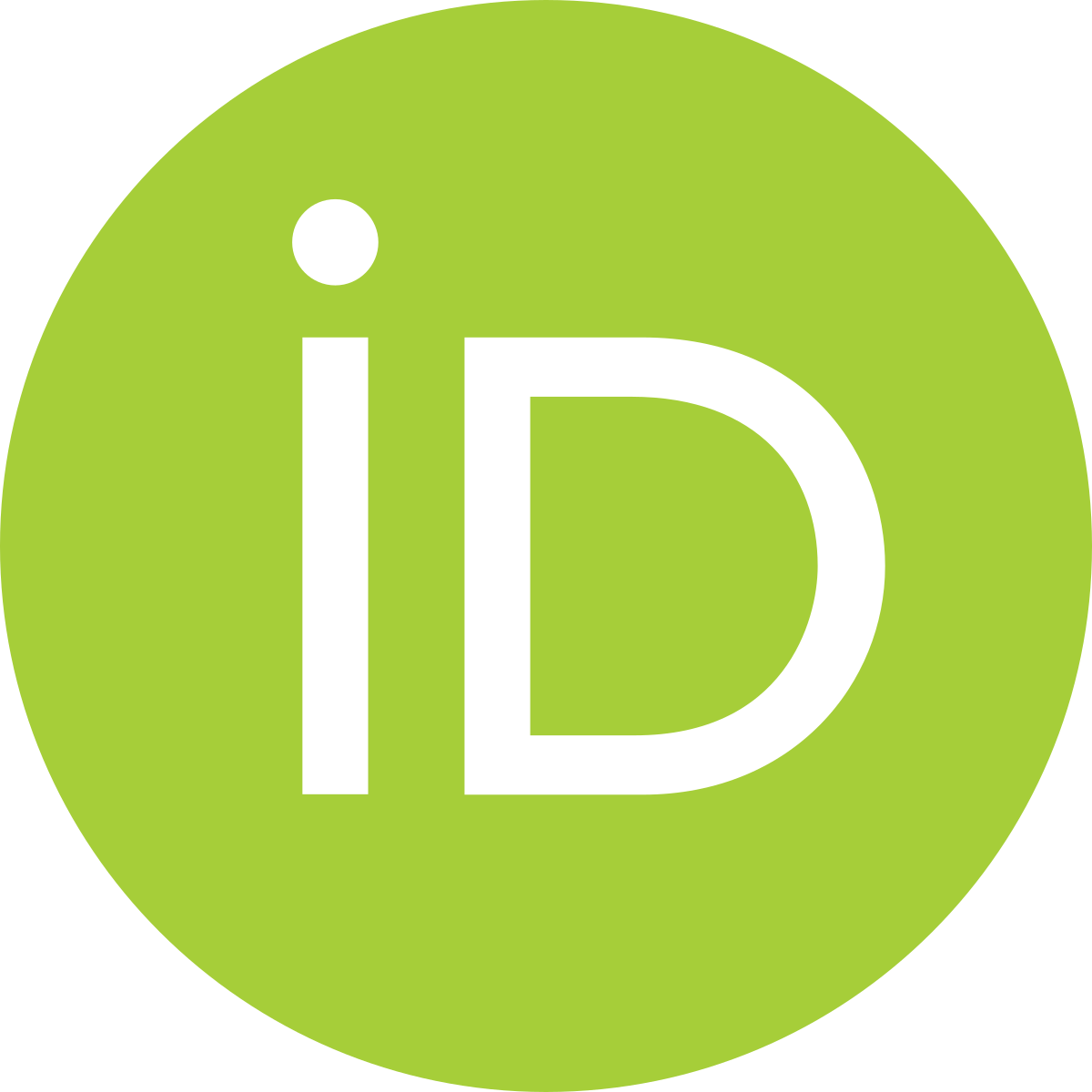}}}
\thanks{These authors contributed equally to this work. Correspondence: andreas.morr@tum.de}
\affiliation{Climate Physics, Max Planck Institute for Meteorology, Germany}

\author{Andreas~Morr\,\href{https://orcid.org/0000-0002-9804-5180}{\includegraphics[width=3.2mm]{orcid.png}}}
\thanks{These authors contributed equally to this work. Correspondence: andreas.morr@tum.de}
\affiliation{Earth System Modelling, School of Engineering \& Design, Technical University of Munich, Germany}
\affiliation{Research Domain IV -- Complexity Science, Potsdam Institute for Climate Impact Research, 14473 Potsdam, Germany}

\author{Klaus~Lehnertz~\href{https://orcid.org/0000-0002-5529-8559}{\includegraphics[width=3.2mm]{orcid.png}}}
\affiliation{Department of Epileptology, University of Bonn Medical Center, 53105~Bonn, Germany}
\affiliation{Helmholtz Institute for Radiation and Nuclear Physics, University of Bonn, 53115~Bonn, Germany}
\affiliation {Interdisciplinary Centre for Complex Systems, University of Bonn, 53175~Bonn, Germany}

\author{Pedro~G.~Lind~\href{https://orcid.org/0000-0002-8176-666X}{\includegraphics[width=3.2mm]{orcid.png}}}
\affiliation{Department of Computer Science, OsloMet -- Oslo Metropolitan University, 0130 Oslo, Norway}
\affiliation{OsloMet Artificial Intelligence lab, OsloMet, 0166 Oslo, Norway}

\author{Niklas~Boers\,\href{https://orcid.org/0000-0002-1239-9034}{\includegraphics[width=3.2mm]{orcid.png}}}
\affiliation{Earth System Modelling, School of Engineering \& Design, Technical University of Munich, Germany}
\affiliation{Research Domain IV -- Complexity Science, Potsdam Institute for Climate Impact Research, 14473 Potsdam, Germany}
\affiliation{Global Systems Institute and Department of Mathematics, University of Exeter, United Kingdom}

\author{Dirk~Witthaut\,\href{https://orcid.org/0000-0002-3623-5341}{\includegraphics[width=3mm]{orcid.png}}}
\affiliation{Forschungszentrum J\"ulich, Institute of Energy and Climate Research (IEK-10), 52428 J\"ulich, Germany}
\affiliation{Institute for Theoretical Physics, University of Cologne, 50937 K\"oln, Germany}

\author{Leonardo~Rydin~Gorj\~ao\,\href{https://orcid.org/0000-0001-5513-0580}{\includegraphics[width=3mm]{orcid.png}}}
\thanks{These authors contributed equally to this work. Correspondence: andreas.morr@tum.de}
\affiliation{Faculty of Science and Technology, Norwegian University of Life Sciences, 1432 Ås, Norway}

\begin{abstract}
Paleoclimate proxy records from Greenland ice cores, archiving e.g. $\delta^{18}$O as a proxy for surface temperature, show that sudden climatic shifts called Dansgaard--Oeschger events (DO) occurred
repeatedly during the last glacial interval. They comprised substantial warming of the Arctic region from cold to milder conditions.
Concomitant abrupt changes in the dust concentrations of the same ice cores suggest that sudden reorganisations of the hemispheric-scale atmospheric circulation have accompanied the warming events.
Genuine bistability of the North Atlantic climate system is commonly hypothesised to explain the existence of stadial (cold) and interstadial (milder) periods in Greenland. 
However, the physical mechanisms that drove abrupt transitions from the stadial to the interstadial state, and more gradual yet still abrupt reverse transitions, remain debated.
Here, we conduct a one-dimensional data-driven analysis of the Greenland temperature and atmospheric circulation proxies under the purview of
stochastic processes.
We take the Kramers--Moyal equation to estimate each proxy's drift and diffusion terms within a Markovian model framework. We then assess noise contributions beyond Gaussian white noise.
The resulting stochastic differential equation (SDE) models feature a monostable drift for the Greenland temperature proxy and a bistable one for the atmospheric circulation proxy.
Indicators of discontinuity in stochastic processes suggest to include higher-order terms of the Kramers--Moyal equation when modelling the Greenland temperature proxy's evolution.
This constitutes a qualitative difference in the characteristics of the two time series, which should be further investigated from the standpoint of climate dynamics.

\end{abstract}

\maketitle

\section{Introduction}
Paleoclimate proxy records provide evidence for past abrupt climate shifts from regional to at least hemispheric scale~\citep[e.g.][and references therein]{Menviel2020, Brovkin2021, Boers2022}. Long-term climate simulations suggest that anthropogenic global warming could trigger structurally similar transitions in several Earth system components in the future, i.e., that these components could `tip' to a qualitatively different state~\citep[e.g.][]{Lenton2008, Lenton2019, Boers2021, ArmstrongMcKay2022, Boulton2022, Wang2023}.
Such catastrophic shifts would have severe consequences on societies and ecosystems and may even unleash feedbacks, further increasing the global mean temperature.
However, the assessment of potentially upcoming tipping points is challenging as the capability of modern complex climate models to simulate climate tipping dynamics is still limited~\citep{Valdes2011, Liu2017, Wang2023}.
In light of this, the study of past abrupt climate shifts may provide insights into the processes involved in climate tipping events.
Furthermore, past events may serve as benchmarks for the performance of fully coupled models in simulating the non-linear and high-dimensional dynamics that could lead to tipping events. In this context, we reassess here two proxy time series from the NGRIP ice core \citep{NGRIP2004}, which feature pronounced imprints of abrupt climatic transitions, by means of the Kramers--Moyal equation.

Agnostic time series models, i.e., models whose dynamics appear to reproduce nature but are not entirely based on physical mechanisms, have played a major role in furthering the debate on climate tipping phenomena \citep[e.g.][]{Riechers2023, Boers2017, Mitsui2017, Kwasniok2013, Lohmann2018a, Dakos2008CSDClimate, Bochow2023SouthAmericanMonsoonTipping}. The ability to produce quantitatively similar dynamical behaviour building only on heuristic physical assumptions facilitates the statistical analysis of tipping phenomena, employing methods of stochastic analysis \citep{Lenton2012CSDClimate, Morr2024RedNoiseCSD, Morr2024KramersMoyalEWS}.
The common concept of a climate tipping element is that of a dynamical system whose current stable equilibrium state is prone to annihilation in a dynamic bifurcation~\citep{Scheffer2009, Ashwin2012, Boers2022}.
This typically involves the reduction of complex, high-dimensional dynamics to just a few (if not one) summary observables that may be modelled in terms of stochastic differential equations (SDEs), i.e., as random dynamical systems.
Therein, the noise term reflects the action of the unresolved dynamics on the summary observable~\citep{Hasselmann1976}.
A common choice is to force the resolved variables with Gaussian white noise, but this approach may be overly simplistic in many situations.
In particular, in the context of climate tipping points a deviation from Gaussian white noise has important implications for the detection of early warning signals and for the probability of premature noise-induced tipping ~\citep{Ditlevsen1999, Lucarini2022, Benson2024alphaStableCSD, Kuehn2022ColourBlind, Morr2024RedNoiseCSD}

Here, we investigate the famous heavy-oxygen $\delta^{18}$O record from the NGRIP ice core~\citep{NGRIP2004}.
The data shows that repeated decadal-scale warming events of regionally up to $16$\,$^{\circ}$C in amplitude, known as Dansgaard--Oeschger events, punctuated the North Atlantic climate throughout the last glacial interval~\citep{Dansgaard1984,Broecker1985,Johnsen1992,Dansgaard1993, Kindler2014}.
The sudden temperature increases were followed by a phase of moderate cooling before the temperatures ultimately relaxed back to colder levels in a second phase of more abrupt cooling.
The two distinct cold and mild regimes are termed stadials and interstadials, respectively. 

In line with the SDE approach outlined above, we regard the $\delta^{18}$O record as a realisation of a one-dimensional Markov process and estimate the corresponding Kramers--Moyal coefficients~\citep{Tabar2019}.
These coefficients are closely related to the Fokker--Planck equation of time-evolving diffusive systems.
Importantly, the analysis suggests including non-zero contributions from higher-order coefficients in the model, which are associated with discontinuous elements in the driving noise~\citep{Anvari2016, Tabar2019}. Such elements can be modelled as Poisson jump processes.
In contrast to the widespread perception of the record as a signature of a genuinely bistable system's trajectory, we find that a monostable underlying deterministic drift is the appropriate choice under the modelling assumption (cf.~Ref.~\citep{Riechers2023} for a two-dimensional analysis.)

We apply the same analysis to the dust record from the same ice core~\citep{Ruth2003, NGRIP2004}.
Concomitant with the $\delta^{18}$O shifts, the dust record exhibits similarly abrupt transitions, which are interpreted as sudden reorganisations of the atmospheric circulation of at least hemispheric scale~\citep{Fuhrer1999, Ruth2003, Ruth2007, Schupbach2018}.
Despite the high degree of correlation between the two records \citep{Boers2017}, the analysis of the dust yields qualitatively different results.
When constructing a time series model, one finds only weak contributions from higher-order Kramers--Moyal coefficients and a bistable deterministic drift function. 

This article is structured as follows:
In Sec.~\ref{sec:2} we briefly introduce the two paleo-climatic proxies that we examine.
Subsequently, in Sec.~\ref{sec:3}, we detail the Kramers--Moyal expansion in one dimension as the prime method to construct time series models including noise and possibly discontinuous elements.
Section~\ref{sec:4} presents the results of this analysis:
Herein, we show the mono- and bistability of the obtained models of the two records and discuss the need to choose a noise model different from Gaussian white noise.
In Sec.~\ref{sec:5} we discuss our findings and relate them to previous work.
Sec.~\ref{sec:6} summarises our key findings and draws conclusions. 

\section{Data and pre-processing}\label{sec:2}
\begin{figure*}[t]
\centering
\includegraphics[width=0.9\linewidth]{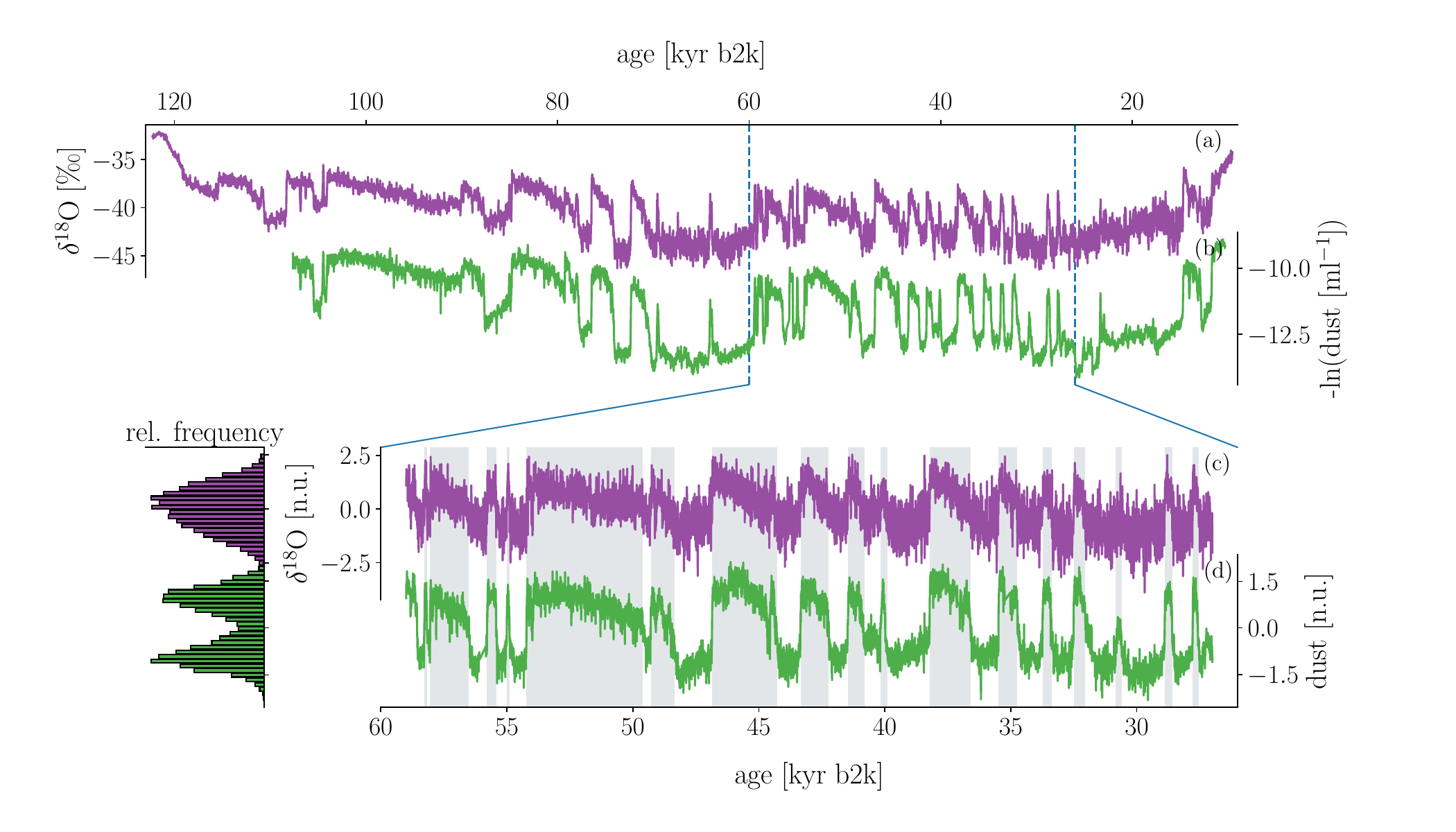}
\vspace{-0.5cm}\caption{Trajectories of the 20-year mean of $\delta^{18}$O (a) and accordingly resampled dust concentrations (b) from the NGRIP ice core in Greenland, from 122\,kyr and 107\,kyr to 10\,kyr before 2000 CE (b2k), respectively~\citep{Ruth2003, Rasmussen2014, Seierstad2014}.
The dust data is given as the negative natural logarithm of the actual dust concentrations, in order to facilitate comparison to the $\delta^{18}$O data.
Panels (c) and (d) show the same proxies but at a higher resolution of 5 years~\citep{NGRIP2004, Gkinis2014, Ruth2003} and over the shorter period from 59 to 27\,kyr~b2k.
The analysis presented in this study was constrained to this segment of the records.
The two proxy time series in (c) and (d) have been detrended by linearly regressing the data against reconstructed global mean surface temperatures~\citep{Snyder2016} and removing the apparent background-temperature-driven slow change. 
The grey shadings mark the Greenland interstadial (GI) intervals according to~\citep{Rasmussen2014}.
All data are shown on the GICC05 chronology~\citep{Vinther2006, Rasmussen2006, Andersen2006, Svensson2008}.
The data were binned to equidistant time resolution from its original 5\,cm depth resolution (see App.~\ref{app:detrending} for further details on the data processing~\citep{Riechers2023}).} \label{fig:1}
\end{figure*}

This work relies on the $\delta^{18}$O and dust concentration records obtained by the North Greenland Ice Core Project (NGRIP)~\citep{Ruth2003, NGRIP2004, Gkinis2014}. From $1404.75$\,m to $2426.00$\,m of depth the joint record is provided at $5$\,cm equidistant resolution. This translates to the time span from $59945$\,yr to $10276$\,yr\,b2k (before 2000 CE) with $\sim5$\,yr resolution for the oldest and sub-annual resolution for the most recent part of the record (Fig.~\ref{fig:1}(a) and (b)). For the analysis, the data was rescaled, binned to an equidistant time axis of 5-year resolution, detrended, and normalised (see Appendix \ref{app:detrending} for details).

The concentration of dust, i.e., the number of particles with a diameter larger than $1$\,$\mu$m per ml, is commonly interpreted as a proxy for the state of the hemispheric atmospheric circulation~\citep[e.g.][]{Fischer2007, Ruth2007, Schupbach2018, Erhardt2019}. In particular, the dust storm activity and dryness over East Asian desserts, the strength and position of the polar jet, and local precipitation patterns govern the emission, transport, and deposition of the dust, respectively~\citep{Fischer2007, Erhardt2019}. Correspondingly, the substantial changes in the dust concentrations at DO events are interpreted as large-scale reorganisations of the Northern Hemisphere's atmospheric circulation. In agreement with a widespread convention, we rescale the dust record by taking the net negative logarithm~\citep[e.g.][]{Ditlevsen1999, Mitsui2017, Boers2017, Riechers2023}. In this form, the dust record exhibits a high degree of correlation with the $\delta^{18}$O record~\citep{Boers2017}. 

In order to reduce the influence of slow changes in the background climate, we restricted the analysis to the period 59--27\,kyr\,b2k and applied further detrending with respect to a Northern Hemisphere temperature reconstruction provided by~\citet{Snyder2016} (see Fig.~\ref{fig:1}(c) and (d) and App.~\ref{app:detrending}). 
The concentration of stable water isotopes expressed as $\delta^{18}$O values in units of permil is a proxy for the site temperature at the time of precipitation~\citep{Jouzel1997, Gkinis2014}.

\section{Methods}\label{sec:3}
Our starting point is a (time-homogeneous) Markov stochastic process $x_t$ of the form
\begin{equation}
  \label{eq:SDE}
  \mathrm{d}x_t = f(x_t) \mathrm{d}t + \sigma(x_t) \mathrm{d}\xi_t, 
\end{equation}
where $\mathrm{d}\xi_t$ denotes an arbitrary uncorrelated stochastic force. The temporal evolution of the associated conditional probability function $p(x,t\!+\!\tau|x'\!,t)$ then follows the Kramers--Moyal (KM) equation~\citep{Kramers1940, Moyal1949, vanKampen1961, Gardiner2009, Risken1996, Tabar2019}:
\begin{equation}\label{eq:KMformal}
    \frac{\partial}{\partial \tau} p(x,t\!+\!\tau|x'\!,t) \!=\! \! \sum_{m=1}^{\infty}\!\!\left ( \!-\frac{\partial}{\partial x} \right )^{\!\!m}\!\!\!D_m(x)\,p(x,t\!+\!\tau|x'\!,t). 
\end{equation}
The KM coefficients $D_m(x)$ are related to the conditional moments $M_m(x,\tau)$ of order $m$ of the stochastic variable $x$ at a time-lag $\tau$ by 
\begin{equation}\label{eq:cond_moments_to_KMcoeffs}
\begin{aligned}
    D_m(x) &= \frac{1}{m!}\lim_{\tau\to 0}  \frac{1}{\tau} M_m(x,\tau)\\
    = \frac{1}{m!}&\lim_{\tau\to 0} \frac{1}{\tau}
    \!\int\! \left( x' \!-\! x_t\right)^m \!p(x',t\!+\!\tau|x,t)\mathrm{d}x'.
\end{aligned}
\end{equation}
In the special case that the stochastic force in Eq.~\eqref{eq:SDE} is given by Gaussian white noise (i.e., it can be expressed by the increments of a Wiener process $W_t$), only the first two terms on the right of Eq.~\eqref{eq:KMformal} contribute and the KM equation reduces to the better-known Fokker--Planck equation \citep{Fokker1913, Fokker1914, Planck1917}.
With $\mathrm{d}\xi_t = \mathrm{d}W_t$, Eq.~\eqref{eq:SDE} becomes the Langevin equation and the resulting process is then referred to as a Langevin process \footnote{There is no agreement on the use of the term Langevin process. Some authors consider Lévy-driven equations as such Langevin equations, others prefer to refer to Langevin processes as those that are solely driven by Gaussian/Brownian noise.}.
For Langevin processes the relation
\begin{equation}\label{eq:Langevin-FP}
  D_1(x) = f(x) \quad \text{and} \quad D_2(x) = \frac{1}{2}\sigma^2(x),
\end{equation}
between the KM coefficients, the drift $f(x)$ and the diffusion $\sigma(x)$, holds in general. 

The other way around, if higher-order moments contribute to the KM equation, the underlying process cannot be a standard Langevin process.
In that case, $\xi_t$ does not correspond to a Wiener process but has instead a more complex form.
However, the first two KM coefficients would still be dominated by the process' drift and diffusion.

While a Langevin process consists, with probability 1, of continuous sample paths \citep[e.g. Theorem 5.1.1 in][] {Arnold1974SDETheoryApplications}, a Markov stochastic process of the form Eq.~\eqref{eq:SDE} generally features discontinuous paths with non-zero probability. Path-wise continuity is only one of many notions of continuity in stochastic processes. Another is the
continuity criterion for Markov processes provided by~\citet{Gardiner2009}, which requires for a process to be continuous that
\begin{equation}\label{eq:Lindberg}
  \begin{split}
  C(x,t,\delta) &= \lim_{\tau\to 0} \frac{1}{\tau} P\left( |x_{t+\tau}- x_t|>\delta \right) \\
  &= \lim_{\tau\to 0} \frac{1}{\tau} \int\limits_{|x'-x|>\delta} p(x', t+ \tau|x,t) dx' \overset{!}{=}0,
  \end{split}
\end{equation}
for all $\delta$, $x$, and $t$. 
The presence of higher-order KM coefficients in the corresponding KM equation is a necessary, yet not sufficient criterion for a given process to be discontinuous under this latter notion.

\subsection{Estimating Kramers--Moyal coefficients}
The central entry point for this work is Eq.~\eqref{eq:cond_moments_to_KMcoeffs}.
It provides a means to estimate the KM coefficients $D_m(x)$ directly from data, i.e., from a recorded realisation of a stochastic process, provided that the following assumptions are fulfilled (to a reasonable degree):
\begin{itemize}
\item[i)] The observed process is a Markov process,
\item[ii)] the process is time-homogeneous, i.e., the dynamics did not change over time,
\item[iii)] the state space is sampled sufficiently densely,
\item[iv)] and the sampling time is short compared to the characteristic time scale of the dynamics.
\end{itemize}
Under these conditions, the evaluation of the conditional statistical moments $M(x,\tau)$ at the shortest available time lag $\Delta t$ given by the sampling rate yields a good estimate for the KM coefficients:
\begin{equation}\label{eq:KMcoeff}
      \hat{D}_{m}(x) = \frac{1}{m!} \frac{1}{\Delta t}\langle\left( x_{t+\Delta t} - x_t\right)^m|_{x_t = x}\rangle \approx D_m(x),
\end{equation}
wherein the ensemble average in Eq.~\eqref{eq:cond_moments_to_KMcoeffs} is replaced by the average over the available data $\langle \cdot \rangle$.
Our numerical implementation of Eq.~\eqref{eq:KMcoeff} is based on the Nadaraya--Watson estimator which is detailed in App.~\ref{App:Bandwidth_selection}.

\subsection{Estimators of discontinuous motion}

Once the KM coefficients are estimated from the data, one can draw inference on the most fitting choice of the noise model $\mathrm{d}\xi_t$.
Vanishing higher-order moments ($m>2$) classify the model as a Langevin process.
In contrast, demonstrable contributions of these moments suggest that the process is best modelled by including noise beyond a Wiener process~\citep[see e.g.][]{vanKampen1961, VanKampen2007, Gardiner2009, Tabar2019,Lin2023}. 

The finite sampling time step $\Delta t$ introduces a bias for the estimators $\hat{D}_{m}(x)$~\citep{Kurth2021JumpDiffusionEpileptic}.
As a consequence, even for a Langevin process the expected values for the higher-order KM estimators differ from zero.
A first pragmatic metric to discern whether a studied process is a Langevin process or not is to evaluate the ratio between the fourth KM coefficient and the second, i.e., $D_4(x)/D_2(x)$.
Small values $\lesssim 0.1$ are typically regarded as a justification for a Langevin description of the data. 
Values $D_4(x)/D_2(x) \gtrsim 0.1$ point to non-diffusive motion (i.e., forcing beyond Gaussian white noise).
This metric offers a first insight into whether a discontinuous noise term $\xi_t$ is needed to model the process~\citep{Gao2016, Yubin2020, Lucarini2022}.

When the Langevin process model is contrasted with a jump-diffusion model of the form~\citep{Tabar2019,Lin2023}
\begin{equation}\label{eq:jump_diff_model}
  \mathrm{d}x_t = f(x_t)\mathrm{d}t + \sigma(x_t) \mathrm{d}W_t + \eta(x_t) \mathrm{d}J^{(\lambda)}_t,
\end{equation}
the assessment can be further refined.
Here, $J^{(\lambda)}_t$ denotes a Poissonian jump process characterised by the rate $\lambda$.
The jump amplitude is determined by the Gaussian stochastic variable $\eta(x)$.
For this specific process model, the KM coefficients read~\citep{Tabar2019}
\begin{equation}\label{eq:jump_diff_KM}
  \begin{aligned}
  D_{1}(x) &= f(x), \\
  D_{2}(x) &= \frac{1}{2}\sigma(x)^2 + \frac{1}{2}\lambda(x) \langle \eta (x)^{2} \rangle,\\ 
  D_{m}(x) &= \frac{1}{m!}\lambda(x) \langle \eta (x)^{m} \rangle, ~\mathrm{for}~m>2,
  \end{aligned}
\end{equation}
where $\langle \cdot\rangle$ expresses the expected value.

Similarly, the bias of the KM estimators defined by Eq.~\eqref{eq:KMcoeff}, when applied to a jump-diffusion process sampled at finite time step $\Delta t$, can be derived analytically.
These considerations offer two additional metrics to distinguish Langevin from jump-diffusion processes, namely the $\Theta$-ratio
\begin{equation}\label{eq:Theta-ratio}
  \Theta(x,\tau) = \frac{3M_2(x,\tau)^2}{ M_4(x,\tau)} \sim\left\{\begin{array}{cl}
        1, & \text{Langevin},  \\
         \frac{1}{\tau}, & \text{jump-diffusion},
  \end{array}\right.
\end{equation}
and the $Q$-ratio~\citep{Lehnertz2018}
\begin{equation}\label{eq:Q-ratio}
  Q(x,\tau) = \frac{M_6(x,\tau)}{5 M_4(x,\tau)} \sim \begin{cases}
        \tau, & \text{Langevin},  \\
         \text{constant},  & \text{jump-diffusion}.
  \end{cases}
\end{equation}
For details on the derivation of these relationships, we refer the interested reader to~\citep{Tabar2019}.
Observing either of the scalings given in Eqs.~\eqref{eq:Theta-ratio} and \eqref{eq:Q-ratio}, respectively, can aid in deciding between employing a Langevin or jump-diffusion model.

For different noise models than the ones introduced above, different scaling behaviours of these ratios with respect to $\tau$ will arise.
In this work, we focus on distinguishing between the Langevin and Poisson jump-diffusion models as two archetypical (dis-)continuous stochastic models.
Observing any other scaling in $Q$ or $\Theta$ may hint at a third model being more appropriate to reproduce the time series dynamics. 
However, in the context of continuous versus discontinuous stochastic models, considering the two discussed models yields essential information.

\begin{figure*}[t]
\includegraphics[width=1\linewidth]{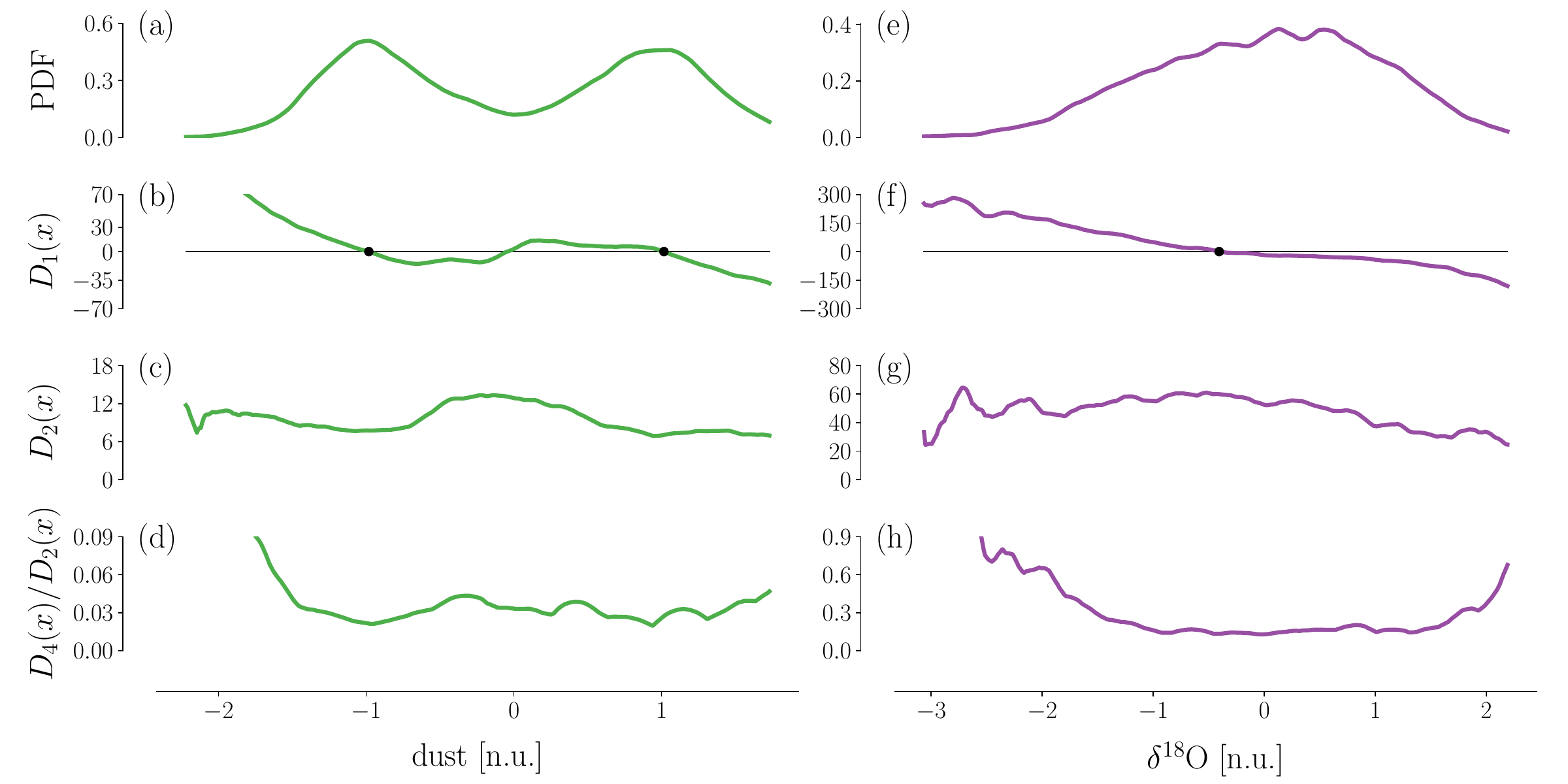}
\vspace{-0.5cm}\caption{The probability density function (PDF) of (a) the dust and (e) $\delta^{18}$O. The non-parametric estimates of the (b, f) first KM coefficient $D_1(x)$ and (c, g) the second KM coefficient $D_2(x)$.
The ratio between the fourth and the second KM coefficient $D_4(x)/D_2(x)$ (d and h).
All KM coefficients are evaluated at the shortest available time step $\Delta t = 5$yr of the time series.
The estimated dust drift is bistable, while that of $\delta^{18}$O is monostable.
The second KM coefficient $D_2(x)$ is relatively constant for both records. The ratio $D_4(x)/D_2(x)$ is small ($\lesssim 0.1$) for the dust record.
Yet, it is non-negligible for $\delta^{18}$O ($\gtrsim 0.3$) in large parts of the state space, suggesting that the driving noise in a stochastic model for these time series should not be exclusively Gaussian white noise.
Details on the choice of kernel and bandwidth used for the KM coefficient estimation, as well as an analysis of the influence of the kernel bandwidth, can be found in App.~\ref{App:Bandwidth_selection}.\label{fig:2}}
\end{figure*}

\begin{figure*}[t]
\includegraphics[width=1.05\linewidth]{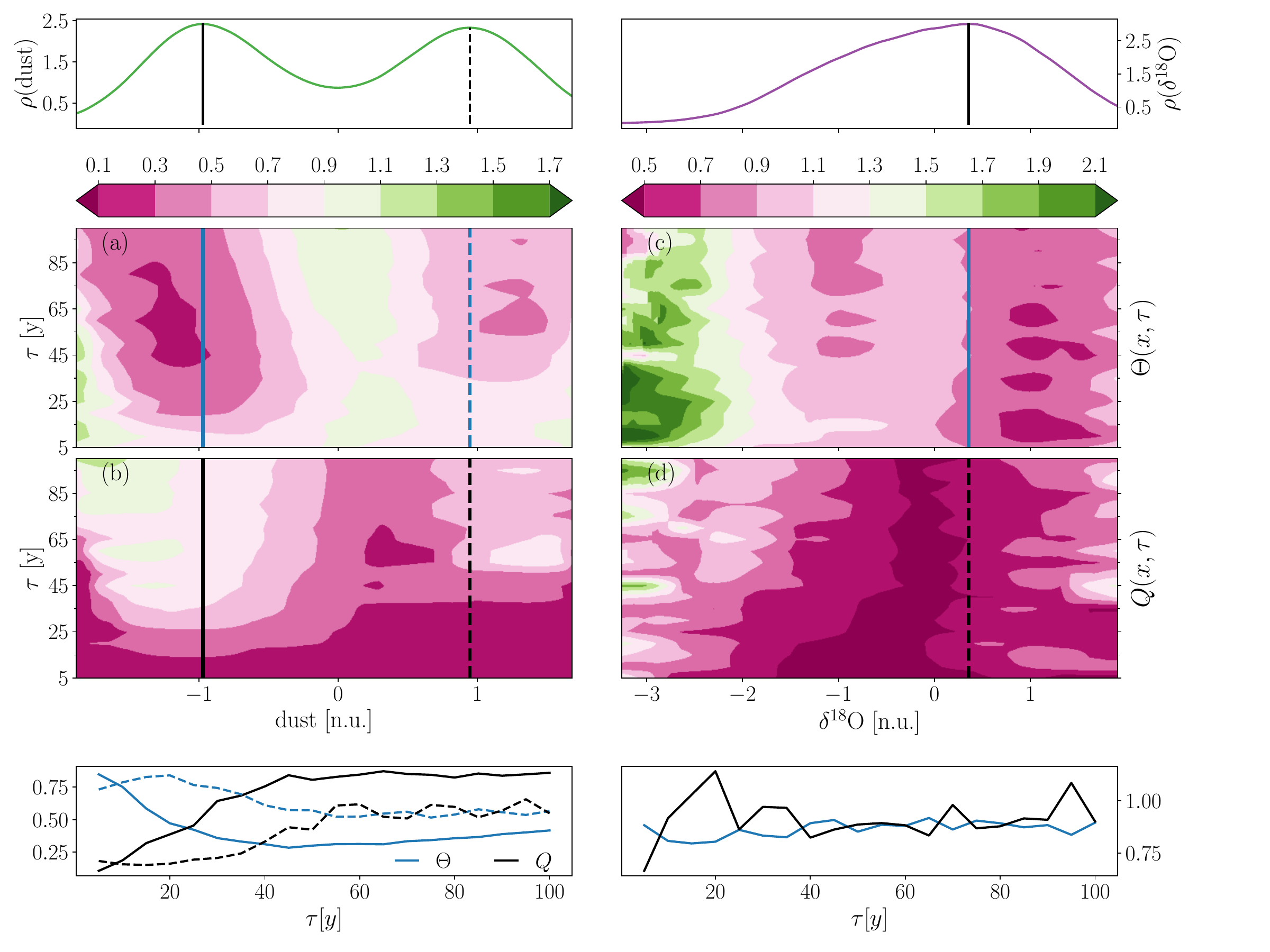}
\vspace{-0.8cm}\caption{The $\Theta(x,\tau)$-ratio and $Q(x,\tau)$-ratio of the dust and the $\delta^{18}$O concentration.
(a) The $\Theta(x,\tau)$-ratio of the dust is not close to $1$ for a large range of $\tau$ and $x$, particularly around the peaks of the bimodal dust distribution.
(b) From $\tau=5$ to roughly $\tau=50$ the $Q(x,\tau)$-ratio increases linearly with $\tau$, consistent with a continuous process, yet for $\tau>50$ the $Q(x,\tau)$-ratio is nearly constant, consistent with a discontinuous process.
(c) On the one hand, the $\Theta(x,\tau)$-ratio of the $\delta^{18}$O points to being different from $1$ along the peak of the distribution, yet not sufficiently conclusive to ascertain if the $\delta^{18}$O is discontinuous.
(d) On the other hand, the $Q(x,\tau)$-ratio is arguably constant over $\tau$, consistent with a discontinuous $\delta^{18}$O.
For visualisation purposes, $Q(x,\tau)$ of $\delta^{18}$O is multiplied by $0.6$ to match the scale of $Q(x,\tau)$ of the dust~\citep{NumPy, SciPy, RydinGorjao2019b, RydinGorjao2023, Matplotlib}.}\label{fig:3}
\end{figure*}

\section{Results}\label{sec:4}
Figure~\ref{fig:2} shows the first and second KM coefficients, and the ratio of the second to the fourth KM coefficients, as estimated from the dust and $\delta^{18}$O time series according to Eq.~\eqref{eq:KMcoeff}.
The corresponding $\Theta$ and $Q$ ratios are presented in Fig.~\ref{fig:3}.

\paragraph{Dust record:}
For the dust, the constructed drift $D_1(x)$ in Fig.~\ref{fig:2}(b) exhibits two separate stable states that match the maxima of the probability density function in Fig.~\ref{fig:2}(a).
The second KM coefficient $D_2(x)$ in Fig.~\ref{fig:2}(c) is approximately constant. The ratio between the fourth and the second KM coefficients in Fig.~\ref{fig:2}(d) is smaller than 0.1 on the entire state space probed by the time series.
For large portions of the dust's state space, we find in Fig.~\ref{fig:3} a decrease of the $\Theta(x,\tau)$ ratio with increasing $\tau$, similar to a $1/\tau$ behaviour. 
This applies, in particular, at the stable equilibria of the drift, where the data availability is the best and our estimation is most robust.
The dust $\Theta(x,\tau)$-ratio is close to $1$ only in a region of its state space where its probability density has a local minimum ($-0.3\lesssim\textrm{dust}\lesssim0.3$).
The corresponding $Q(x,\tau)$-ratio shows quite a distinct linear increase with increasing $\tau$ -- at least for small values of $\tau$.
For larger values of $\tau$, $Q(x,\tau)$ is constant.

\paragraph{$\delta^{18}O$ record:}
In the case of $\delta^{18}$O, the drift has only one zero-crossing in agreement with the unimodal distribution of the data.
With respect to the normalised units, the first and second KM coefficients of $\delta^{18}$O exceed their counterparts for dust by factors of approximately $4$ and $10$, respectively.
This indicates that $\delta^{18}$O was subjected to stronger noise while simultaneously stronger deterministic forces acted on the variable.
Finally, the ratio $D_4(x)/D_2(x)\gtrsim0.3$ is $10$ times larger for $\delta^{18}$O than for the dust.
The $\delta^{18}$O record exhibits a mostly constant $\Theta(x,\tau)$-ratio with respect to $\tau$, as seen in Fig.~\ref{fig:3}.
It is slightly below but still close to $1$ for large parts of the state space. 
The corresponding $Q(x,\tau)$-ratio is likewise constant ($\approx 1$) with respect to $\tau$, with variations in both directions.

\section{Discussion}\label{sec:5}
The assessment of the KM coefficients and the scaling of $\Theta$ and $Q$ ratios from the dust and the $\delta^{18}$O records provides some insight into how to best model the proxy time series within the framework of one-dimensional stochastic processes.

For the dust, we find bistability of the estimated model's drift.
The small $D_4(x)/D_2(x)$ ratio and the linear increase of the $Q(x,\tau)$ with increasing $\tau$ indicate that this process can, in fact, be modelled as a Langevin process.
Only the assessment of the $\Theta(x,\tau)$ ratio calls this conclusion into question.
For a Langevin process, this ratio should be equal to one, but we observe a $1/\tau$-like scaling for small values of $\tau$, in line with an underlying jump-diffusion process. 

We note that a simple Langevin model with a bistable drift and purely diffusive noise can produce the regime shifts observed in the dust record. However, such a model is unlikely to reproduce the asymmetric shape of the interstadial phases evident in the record. 

For the $\delta^{18}$O record, the results are exactly the opposite.
The constructed drift function exhibits only a single stable equilibrium.
The observed quantities $D_4(x)/D_2(x)$ and $Q(x,\tau)$ provide evidence for relevant contributions from higher-order KM coefficients. The $\Theta(x,\tau)$ ratio, however, is close to one in agreement with a Langevin model.
A Langevin model together with the evidenced single equilibrium of the drift function clearly fails to explain the two regimes of the $\delta^{18}$O record, and the apparent time asymmetry.
Taken together, we conclude that the evidence speaks in favour of introducing discontinuities to the driving noise model rather than against it.
Complex noise, i.e., noise beyond a Wiener process, could indeed be a way to reproduce time series with two regimes in the presence of a single equilibrium and time asymmetry~\citep{Chechkin2003, Chechkin2004, Metzler2004, Yang2020}. 

Given the high degree of visual similarity between the dust and the $\delta^{18}$O records, the differences in the reconstructed potentials and the ratio between the fourth and the second KM coefficient are remarkable.
This accentuates the need for careful statistical analysis when devising time series models for non-linear systems with abrupt transitions.

Adopting a generalised Langevin equation with a bistable drift term, Ditlevsen~\citep{Ditlevsen1999} showed that the noise in the calcium concentration record from the GRIP ice core can be modelled with an $\alpha$-stable component.
Calcium concentrations are typically considered equivalent to dust concentrations (cf.~\citep{Fuhrer1993, Ruth2002, Ruth2003, Fischer2007}).
We cannot directly assess the presence of $\alpha$-stable noise in the NGRIP dust record. This is because noise models with infinite statistical moments, which can be found in $\alpha$-stable distributions, are inherently incompatible with the Kramers--Moyal framework.
Yet, our results corroborate the notion that Greenland ice core records bear the signature of non-Gaussian noise, though in our analysis this arises primarily for the $\delta^{18}$O record.
Related to this, \citet{Gottwald2020} recently formulated a conceptual model of DO events wherein $\alpha$-stable noise plays a central role as an event trigger, later extended by~\citet{Riechers2023}. From the perspective of theoretical stochastic modelling, it is worth noting that the $\alpha$-stable noise model leads to a path-wise continuous process, in contrast to the Poisson jump-diffusion model discussed in this work.
Employing the continuity notion of Eq.~\eqref{eq:Lindberg}, however, both of these models would be considered discontinuous.

We have to state that the interpretation of higher-order KM coefficients is not straightforward and depends on the exact choice of the stochastic model.
A direct causal relation between the DO events and discontinuous noise cannot be inferred without further ado within this study, but the role of discontinuities in the proxy records merits further investigation.
It has been observed in complex model simulations that (stochastic) atmospheric anomalies can indeed drive regime changes in the North Atlantic region~\citep{Drijfhout2013, Kleppin2015}. 
Together with the apparent aptitude of non-Gaussian noise models for Greenland temperature and Northern Hemisphere atmospheric circulation proxies, this motivates further research on the effect that non-Gaussian noise could have on climate tipping elements in present-day climate. 

If both Greenland temperatures and the state of the Northern Hemisphere atmospheric circulation were subject to non-Gaussian noise, and if indeed pulses of this noise triggered transitions between stadial and interstadial regimes, this would have important implications for our conception of stability of certain climate tipping elements.
The possibility that climate tipping elements are nowadays likewise subject to non-Gaussian stochastic forcing warrants more attention.  

\section{Conclusion}\label{sec:6}
In this work, we presented a data-driven analysis of the $\delta^{18}$O and dust concentration records from the NGRIP ice core, based on the Kramers--Moyal equation.
This equation generalises the Fokker--Planck equation by allowing for arbitrarily complex uncorrelated driving noise $\mathrm{d}\xi_t$.
In particular, such noise may result in a discontinuous process.

The estimation of the KM coefficients yielded a monostable drift for the isolated $\delta^{18}$O record and a bistable one for the dust.
The analysis of the resulting agnostic time series models does not allow for conclusions about the dynamical stability of the actual physical processes. It is, however, notable that these findings are inconsistent with the hypothesis that past Greenland temperatures were governed by intrinsically bistable dynamics \citep{Livina2010, Kwasniok2013}. For the atmospheric circulation there arises no such inconsistency. 

We found that stochastic forcing should include terms beyond Gaussian white noise when modelling the $\delta^{18}$O record. 
This renders the Langevin approach insufficient to accurately reproduce the time series characteristics, drawing attention towards including discontinuous elements. For the dust record, similar indications could be found, though these have not been as convincing.

In physical terms, complex noise could have played a central role in the emergence of DO events.
Our analysis does not provide direct evidence for a causal relation between discontinuous driving noise and the regime switches of the North Atlantic region's climate during the last glacial.
Yet, it motivates further exploration of this issue along the lines of \citet{Gottwald2020} and \citet{Riechers2023}.
The possibility that climate tipping elements are subject to non-Gaussian noise in today's climate should receive greater consideration.
The corresponding implications on the stability of these elements and the ability to detect early warning signals should be investigated.

\section*{Acknowledgements}
We thank Peter Ditlevsen for the conversations and early remarks on this manuscript that helped improve its content.
LRG and DW gratefully acknowledge support from the Helmholtz Association via the grant \textit{Uncertainty Quantification -- From Data to Reliable Knowledge (UQ)} with grant agreement no.~ZT-I-0029.
This work was performed by LRG as part of the Helmholtz School for Data Science in Life, Earth and Energy (HDS-LEE). 
NB acknowledges funding from the Volkswagen Foundation.
Funded by the Deutsche Forschungsgemeinschaft (DFG, German Research Foundation), grant no.~491111487.
This is a ClimTip contribution. The ClimTip project has received funding from the European Union's Horizon Europe research and innovation programme under grant agreement No.~101137601. Views and opinions expressed are however those of the author(s) only and do not necessarily reflect those of the European Union or the European Climate, Infrastructure, and Environment Executive Agency (CINEA). Neither the European Union nor the granting authority can be held responsible for them.

\section*{Supplementary information}

The code used for this study will be made available by the authors upon request.

All ice core data can be obtained from the website of the Niels Bohr Institute of the University of Copenhagen (\url{https://www.iceandclimate.nbi.ku.dk/data/}); the detailed links are indicated below.
The original measurements of $\delta^{18}$O and dust concentrations go back to~\citep{NGRIP2004} and~\citep{Ruth2003}, respectively. 
The 5\,cm resolution $\delta^{18}$O and dust concentration data together with corresponding GICC05 ages used for this study can be downloaded from \url{https://www.iceandclimate.nbi.ku.dk/data/NGRIP_d18O_and_dust_5cm.xls} (accessed 2025-01-07). 
The $\delta^{18}$O data shown in Fig.~\ref{fig:1} with 20\,yr resolution that cover the period 122-10 kyr b2k are available from \url{https://www.iceandclimate.nbi.ku.dk/data/GICC05modelext_GRIP_and_GISP2_and_resampled_data_series_Seierstad_et_al._2014_version_10Dec2014-2.xlsx} (accessed 2025-01-07) and were published in conjunction with the work by~\citet{Rasmussen2014} and~\citet{Seierstad2014}. 
The corresponding dust data, also shown in Fig.~\ref{fig:1} covering the period 108--10\,kyr~b2k, can be retrieved from \url{https://www.iceandclimate.nbi.ku.dk/data/NGRIP_dust_on_GICC05_20y_december2014.txt} (accessed 2025-01-07). 
The global average surface temperature reconstructions provided by \citet{Snyder2016} and used here for the detrending were retrieve  from \url{https://static-content.springer.com/esm/art\%3A10.1038\%2Fnature19798/MediaObjects/41586_2016_BFnature19798_MOESM258_ESM.xlsx} (accessed 2025-01-07)

\bibliography{bib}

\appendix

\section{Data detrending}\label{app:detrending}

\begin{figure*}[t]
\includegraphics[width=1\linewidth]{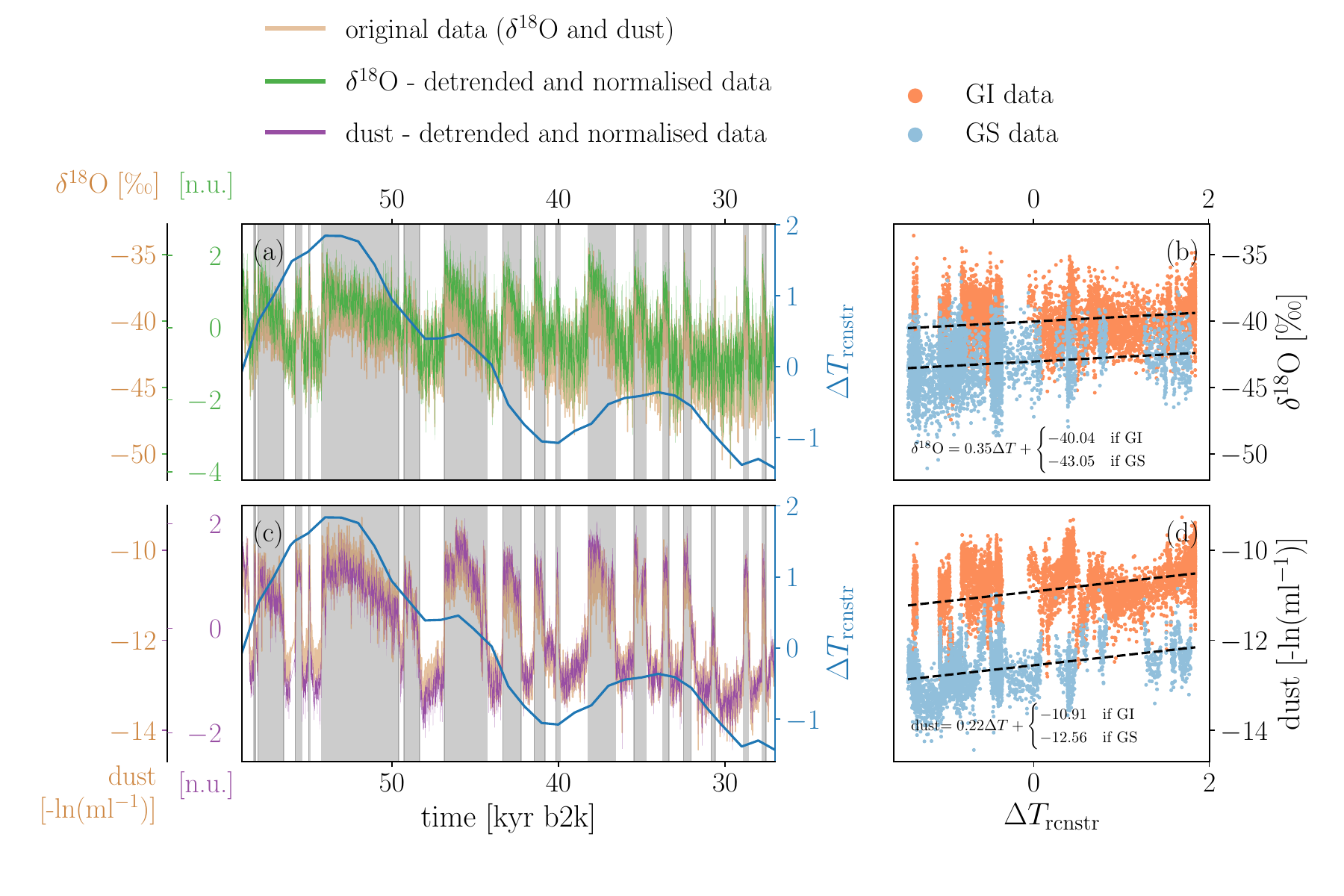}
\vspace{-1cm}\caption{Removal of a linear trend in the NGRIP $\delta^{18}$O and dust time series~\citep{NGRIP2004} with respect to a global average surface temperature reconstruction~\citep{Snyder2016}.
In panel (a) both original $\delta^{18}$O (light brown) as well as detrended and normalised (purple) time series are shown.
Likewise for the dust record in panel (b) (light brown and green, respectively).
The background temperature is given in anomalies with respect to the mean over the investigated period (blue).
Panels (c) and (d) show scatter plots of the original $\delta^{18}$O and dust data with respect to temporarily corresponding temperature anomalies, respectively.
Data from interstadials (stadials) is shown in orange (light blue). The black dashed lines correspond to the fitting scheme that uses a single slope but two different offsets to separately fit the stadial and interstadial data.}\label{fig:detrending}
\end{figure*}

As mentioned in Sec.~\ref{sec:2}, this study focuses on the period 59--27\,kyr~b2k.
Detrending of the data is necessary to ensure that the time series are time-homogeneous stationary processes, which is an underlying assumption for the Kramers--Moyal analysis performed in our investigation.
To compensate for the influence of the background climate on the climate proxy records of dust and $\delta^{18}$O, we remove a linear drift with respect to reconstructed global average surface temperatures~\citep{Snyder2016} from both time series.
Figure~\ref{fig:detrending} illustrates the detrending scheme. 
Due to the two-regime nature of the time series, a simple linear regression would overestimate the temperature dependencies. 
Instead, we separate the data from Greenland stadials (GS) and Greenland interstadials (GI) and then minimise the expression
\begin{equation}
    R^2 \!=\! \sum_{i=1}^{N} \left(\!X_{t_i} - a_X \Delta T(t_i) - \begin{cases} b_{\text{GI}}, ~ \mathrm{if}~t_i \in \text{GI}\\
    b_{\text{GS}}, ~ \mathrm{if}~t_i \in \text{GS}\!\!\!\end{cases}\!\!\!\!\right)^{\!\!2}, 
\end{equation}
with $X$ either dust or $\delta^{18}$O and with respect to the parameters $a_X$, $b_{\text{GI}}$, and $b_{\text{GS}}$.
For a given time $t_i \in \text{GS\;(GI)}$ indicates that $t_i$ falls into a stadial (interstadial) period.
The resulting $a_X$ is used to detrend the original data with respect to the temperature. 
The detrended data are subsequently normalised by subtraction of their mean and division by their standard deviation. 

\section{Nadaraya--Watson estimator of the Kramers--Moyal coefficients and bandwidth selection}\label{App:Bandwidth_selection}

\begin{figure*}[t]
\includegraphics[width=1\linewidth]{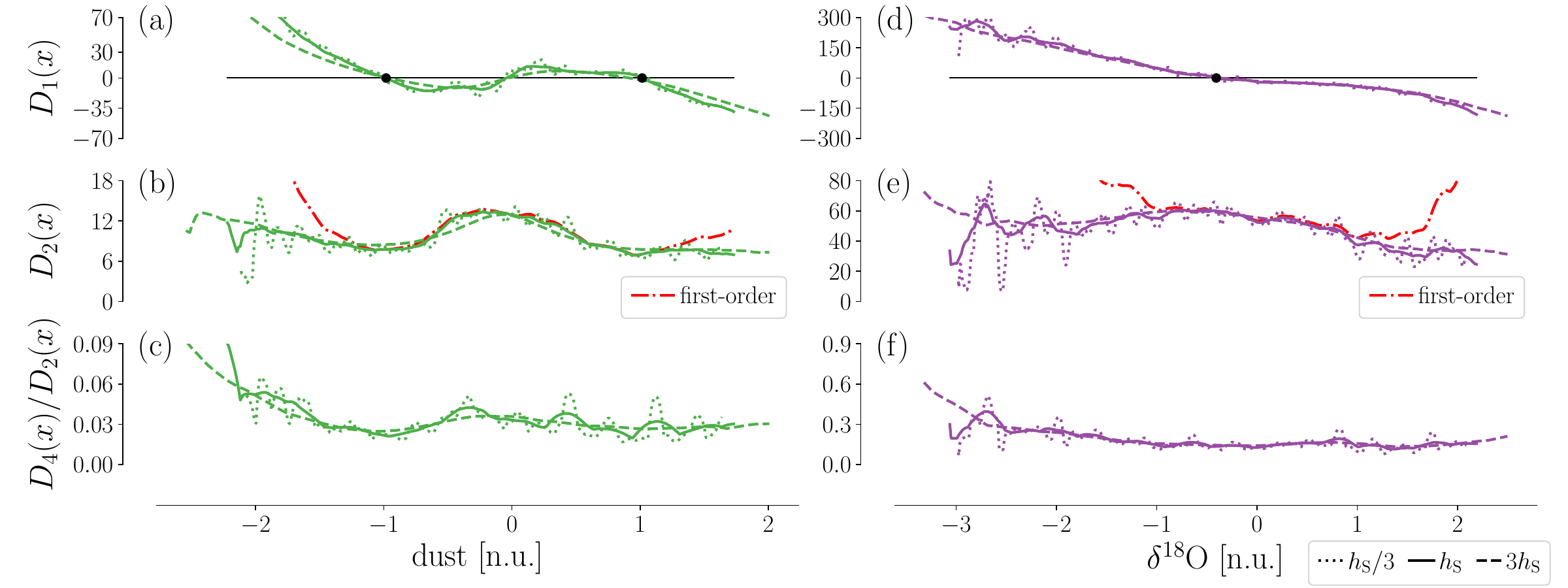}
\vspace{-0.5cm}\caption{The effect of the bandwidth selection $h_\mathrm{S}$ on the KM estimations, in similar fashion to Fig.~\ref{fig:2}.
The non-parametric estimates of the first KM coefficient $D_1(x)$, the second KM coefficient $D_2(x)$, and the ratio of the fourth to the second KM coefficient $D_4(x)/D_2(x)$.
Left column for dust, right column for $\delta^{18}$O.
Three bandwidths are used for the Nadaraya--Watson kernel-density estimator: the optimal Silverman's rule-of-thumb $h_\mathrm{S}$, three times $h_\mathrm{S}$, and one-third $h_\mathrm{S}$.
The Nadaraya--Watson kernel-density estimator's bandwidths $h_\mathrm{S}$ for $\delta^{18}$O is $0.131$ and for dust $0.103$.
In all cases, the interpretation of the estimator remains the same: bistability in the dust, mono-stability in the $\delta^{18}$O. 
In (b) and (e) the first-order estimator for the second KM coefficient $D_2(x)$ are included, i.e., without corrective terms.}\label{fig:6}
\end{figure*}

In order to carry out the estimation in Eq.~\eqref{eq:KMcoeff} we map each data point in the corresponding state space to a kernel density and then take a weighted average over all data points 
\begin{equation}
\begin{aligned}
    D_m(x) &\sim \frac{1}{m!}\frac{1}{\Delta t}\langle (x_{t+\Delta t} - x_t)^m|_{x_t = x} \rangle \\
    & \sim \frac{1}{m!}\frac{1}{\Delta t}\frac{1}{N}\sum_{i=1}^{N-1} K(x - x_i)(x_{i+1}-x_{i})^m.
\end{aligned}
\end{equation}
Similarly to selecting the number of bins in a histogram, for a kernel-density estimation, we select both a kernel and a bandwidth~\citep{Nadaraya1964, Watson1964, Lamouroux2009}.
The kernel is a function $K(x)$ for the estimator $\widehat{f}_h(x)$, where $h$ is the bandwidth at a point $x$, following
\begin{equation}
    \widehat{f}_h(x) = \frac{1}{nh} \sum_{i=1}^n K\left(\frac{x-x_i}{h}\right)
\end{equation}
for a collection $\{x_i\}$ of $n$ random variables.
The kernel $K(x)$ is normalisable ${\int\!K(x)\mathrm{d} x = 1}$ and has a bandwidth $h$, such that $K(x) = K(x/h)/h$~\citep{RydinGorjao2019, Tabar2019, Davis2022}.
The bandwidth $h$ is equivalent to the selection of the number of bins, except that binning in a histogram is always `placing numbers into non-overlapping boxes'.
We use an Epanechnikov kernel
\begin{equation}
    K(x) = \frac{3}{4}(1-x^2), ~\mathrm{with}~\mathrm{support}~|x|<1.
\end{equation}
which has a compact bounded support, but other kernels are available, with different supports~\citep{Epanechnikov1967}.
The selection of an appropriate bandwidth $h$ follows Silverman's rule-of-thumb~\citep{Silverman1998}, given by
\begin{equation}
    h_\mathrm{S} = \left(\frac{4\hat{\sigma}^5}{3n}\right)^{\frac{1}{5}},
\end{equation}
where $\sigma^2$ is the variance of the time series.
In Fig.~\ref{fig:6} three different bandwidths are used to evaluate the various KM coefficients, as given in Fig.~\ref{fig:2}. 
The bandwidths are the optimal bandwidth given by Silverman's rule-of-thumb $h_\mathrm{S}$, three times $h_\mathrm{S}$, and one-third $h_\mathrm{S}$.

Note that regardless of the choice of bandwidth, the mono-stability of the $\delta^{18}$O model is preserved, as is the bistability of the dust concentration model.

\section{Understanding continuity and discontinuity in stochastic processes}\label{app:discontinuous}

The understanding of \textit{continuity} and \textit{discontinuity} can sometimes be unclear when dealing with time series data.
We turn to Lindeberg's continuity condition $C(x,t,\delta)$ for a Markov process~\citep{Lehnertz2018, Tabar2019}, which states that a process $x_t$ is continuous if 
\begin{equation}
\begin{aligned}
    C(x,t,\delta) &= \lim\limits_{\tau\to 0}\frac{\mathrm{Prob}\left[\left.|x_{t+\tau}-x_t|>\delta\right|_{x_t=x}\right]}{\tau}\\
    &= \lim\limits_{\tau\to 0}\frac{\int_{|x'-x|>\delta}p(x',t+\tau|x,t)\mathrm{d} x'}{\tau} \\
    &= 0.
\end{aligned}
\end{equation}
In words, this means that the probability of a particle deviating from a reference position more than $\delta$ in a time interval $\tau$ decreases faster than  inearly with $\tau$.
Direct proof is easily obtained for some particular processes.
For example, for a Brownian motion we obtain, as expected, $C(x,t,\delta)=0$ (see~\citep{Tabar2019}, Eq.~(4.5) for a derivation).
In a similar fashion, \citet{Tabar2019} also shows two examples where $C(x,t,\delta)>0$.
These are the Cauchy process (which is the special case of an $\alpha$-stable Lévy-driven Langevin process with $\alpha=1$) and processes with Poissonian jumps (see Eq.~(4.6) and Eq.~(11.19)).
Both examples are discontinuous processes by this definition. 

For discontinuous processes in our KM setting we can derive a relation similar in form to Lindeberg's continuity condition, namely $C(x,t,\delta)\leq\frac{\bar{M}_m(x)}{\delta^m}$.
We follow almost verbatim the derivation by Tabar, S11.2~\citep{Tabar2019}.
Consider the $m$-th order conditional moment of the \textit{absolute value} of the increment
\begin{equation}
\begin{aligned}
    \langle |x_{t+\tau}-x_t|^m&|_{x_t = x} \rangle = \langle |x'-x|^m|_{x_t = x}\rangle\\
    &=\int\limits_{-\infty}^{\infty}|x'-x|^m p(x',t+\tau|x,t)\mathrm{d} x'\\
    &\geq \mkern-18mu\int\limits_{|x'-x|>\delta}\mkern-18mu|x'-x|^m p(x',t+\tau|x,t)\mathrm{d} x',\\
\end{aligned}
\end{equation}
where we disregarded the integration over the interval $[x-\delta,x+\delta]$, $\delta$ being a small value.
Using further that $|x'-x|^m>\delta^m$, we get
\begin{equation}
    \langle |x'-x|^m|_{x_t = x}\rangle  \geq \delta^m\mkern-18mu\int\limits_{|x'-x|>\delta}\mkern-18mu p(x',t+\tau|x,t)\mathrm{d} x'.
\end{equation}
Dividing both sides by $\tau$ and taking the limit $\tau\to 0$, we obtain
\begin{equation}
    \lim\limits_{\tau\to0}\frac{1}{\tau}\delta^m\mkern-22mu\int\limits_{|x'-x|>\delta} \mkern-18mu p(x',t+\tau|x,t)\mathrm{d} x' \leq \lim\limits_{\tau\to0}\frac{1}{\tau} \langle |x'-x|^m|_{x_t = x}\rangle,
\end{equation}
where we recognise the form of Lindeberg's continuity condition as
\begin{equation}\label{eq:lindeberg}
	C(x,t,\delta)\leq\frac{\bar{M}_m(x)}{\delta^m},
\end{equation}
with
\begin{equation}
    \bar{M}_m(x) = \lim\limits_{\tau\to 0}\frac{1}{\tau} \langle |x'-x|^m|_{x_t = x}\rangle,
\end{equation}
noting the absolute value in contrast with Eq.~\eqref{eq:cond_moments_to_KMcoeffs}.
Although the relation comprises only an upper bound for $C(x,t,\delta)$, it yields a convincing argument for the role of the higher-order KM coefficients and their relation with discontinuity.

Note that having any vanishing KM coefficient of order $m>2$ is sufficient for the process to be continuous.
For the case of non-vanishing KM coefficients of higher order, Pawula's theorem~\citep{Pawula1967a, Pawula1967b, Risken1996} implies that all KM coefficients exist.
It is reasonable to expect Lindeberg's continuity condition will \textit{not} be obeyed for at least one order $m$ (note the left-hand side of Eq.~\eqref{eq:lindeberg} does not depend on $m$).
Consequently, higher-order KM coefficients relate to discontinuous trajectories.
This, however, does not imply that the Kramers--Moyal equation is necessarily the only model -- or even the best model -- to describe discontinuous processes~\citep[cf.][]{VanKampen2007, Gardiner2009}.

\end{document}